# Enhanced Robot Speech Recognition Using Biomimetic Binaural Sound Source Localization

Jorge Dávila-Chacón, Jindong Liu, *Member, IEEE,* and Stefan Wermter

*Abstract*— Inspired by the behavior of humans talking in noisy environments, we propose an embodied embedded cognition approach to improve automatic speech recognition (ASR) systems for robots in challenging environments, such as with ego noise, using binaural sound source localization (SSL). The approach is verified by measuring the impact of SSL with a humanoid robot head on the performance of an ASR system. More specifically, a robot orients itself toward the angle where the signal-to-noise ratio (SNR) of speech is maximized for one microphone before doing an ASR task. First, a spiking neural network inspired by the midbrain auditory system based on our previous work is applied to calculate the sound signal angle. Then, a feedforward neural network is used to handle high levels of ego noise and reverberation in the signal. Finally, the sound signal is fed into an ASR system. For ASR, we use a system developed by our group and compare its performance with and without the support from SSL. We test our SSL and ASR systems on two humanoid platforms with different structural and material properties. With our approach we halve the sentence error rate with respect to the common downmixing of both channels. Surprisingly, the ASR performance is more than two times better when the angle between the humanoid head and the sound source allows sound waves to be reflected most intensely from the pinna to the ear microphone, rather than when sound waves arrive perpendicularly to the membrane.

## I. Introduction

HUMANS routinely perform complex behaviors that are important for surviving in dynamic environments. This range of conducts is supported by an internal representation of the world acquired through our senses. Even though the information we receive is subject to noise from several sources, integration of different sensory modalities can provide the necessary redundancy to perceive the environment with consistency. In the case of auditory perception, our nervous system is capable of extracting different kinds of information contained in sound. We perform low-level processing of sound in the first layers of our auditory pathway. These initial stages allow us to segregate individual sound sources from a noisy background, localize them in space, and detect their motion patterns [1, Ch. 5]. Afterwards, in latter stages of auditory processing, we are able to accomplish high-level auditory tasks such as understanding natural language [1, Ch. 4].

Although the neurophysiology of the mammalian auditory pathway has been extensively studied in the past decades, few research has been done about sound source localization (SSL) and automatic speech recognition (ASR) inside the framework of embodied cognition [2]. Particularly, further research is needed to integrate the cues used by human listeners that are not present in traditional ASR methods [3], e.g., emergent language segmentation and multimodal integration. Nevertheless, ample literary resources already provide a solid basis for bioinspired technological applications [1], [4]–[6].

Our objective is to understand the influence of human physiognomy on SSL and ASR. If, from a Human–Robot Interaction point of view, a human is the best interface for another human [7], we should exploit the computational advantages that physiognomy brings in *for free*. For this reason, we use the iCub humanoid to measure the influence that the body has on our models of the auditory system. Afterwards, we compare the results obtained with a dummy head designed for binaural recordings.

Once the anthropomorphic geometry of the robot produces the spatial cues, we want to find a principled method to integrate them, as they are complementary sensory modalities. Recent work from our group shows that neural methods can achieve near-optimal integration of multiple sensory modalities [8], so we integrate the spatial cues following the same principles. Eventually, this should lead to the use of robotic SSL for improving the accuracy of ASR systems. A common challenge with robotic platforms is the presence of noise produced by the robot's cooling system. Hence, it is also important to develop a system that can overcome interference of such *ego noise* near the microphones.

In order to construct a bioinspired model for SSL, it is necessary to examine the current theories about the neural encoding of auditory spatial cues. More specifically, it is important to understand how our nervous system represents and integrates such cues along the auditory pathway. In Section I-A, we further describe the neuroanatomy and neurophysiology relevant for SSL.

Manuscript received June 19, 2016; revised April 26, 2017, July 25, 2017, February 7, 2018, and April 16, 2018; accepted April 22, 2018. Date of publication June 4, 2018; date of current version December 19, 2018. This work was supported in part by the DFG German Research Foundation–International Research Training Group Cross-Modal Interaction in Natural and Artificial Cognitive Systems under Grant 1247 and in part by DFG through the Cross-Modal Learning Project under Grant TRR 169 *(Corresponding authors: Jorge Dávila-Chacón; Jindong Liu.)*

J. Dávila-Chacón and S. Wermter are with the Knowledge Technology Group, Department of Informatics, University of Hamburg, Vogt-Kölln-Straße 30, 22527 Hamburg, Germany (e-mail: davila@informatik.uni-hamburg.de).

J. Liu is with the Department of Computing, Imperial College London, South Kensington Campus, London SW7 2AZ, U.K. (e-mail: j.liu@imperial.ac.uk).

Color versions of one or more of the figures in this paper are available online at http://ieeexplore.ieee.org.

Digital Object Identifier 10.1109/TNNLS.2018.2830119

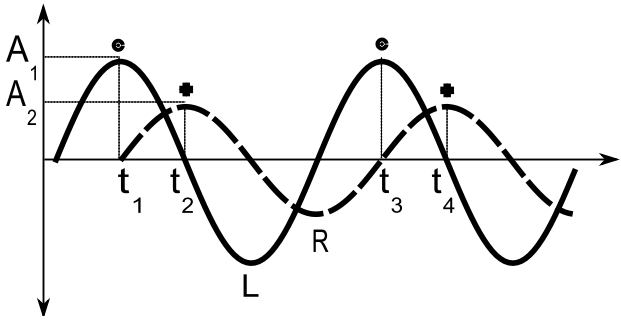

Fig. 1. Waves represent the vibrations in the left (L) and right (R) basilar membranes, at sections resonant to a given sound frequency component $f$. The auditory system is known to compare the timing of neural spikes when the time delay between them is less than half a period [1, Ch. 5.3.3]. Therefore, our MSO model considers the time difference $\Delta t$ between $t_1$ and $t_2$ for the computation of ITDs, but not the $\Delta t$ between $t_2$ and $t_3$. ILDs are computed in our LSO model as the logarithmic ratio of the vibration amplitudes at $t_1$ and $t_2$ as $\log(A_1/A_2)$.

### A. Neural Correlates of Sound Source Localization

When sound waves approach our body, they are affected by the absorption and reflection of our torso, head, and pinnae. This interaction modifies the frequency spectrum of sound reaching our ear canal in different ways, depending on the spatial location of the sound source around our body. Once the sound waves reach our inner ear, they produce vibrations inside the cochlea. The information contained in these vibration patterns is then encoded by the organ of Corti, where mechanical vibrations in the basilar membrane are transduced into neural spikes. Afterward, these spikes are delivered through the auditory nerve to the cochlear nucleus, a relay station that forward information to the medial superior olive (MSO) and to the lateral superior olive (LSO). The MSO and LSO are of our particular interest because they extract interaural time differences (ITDs) and interaural level differences (ILDs) respectively. The waves shown in Fig. 1 represent vibrations in the left (L) and right (R) basilar membranes at a section resonant to a given sound frequency component $f$. The markers above the maximum amplitudes of the waves represent the point in time with the maximum probability of a neural spike to be produced by the hair cells in the organ of Corti.

The MSO performs the task of a coincidence detector, where different neurons represent spatially different ITDs [9]. Neurons in the MSO encode ITDs more effectively from the low-frequency components of sound. This representation can be achieved by different delay mechanisms, such as different thicknesses of the axon myelin sheaths, or different axon lengths from the excitatory neurons in the ipsilateral and contralateral cochlear nucleus [10]. The principle behind these mechanisms is represented in Fig. 2. In the case of level differences, different neurons in the LSO represent spatially different ILDs. Due to the shadowing effect of the head, the LSO encodes ILDs more effectively from the high-frequency components of sound [11]. The mechanism underlying the extraction of ILDs is less clear than the one of ITDs. Nevertheless, it is known that LSO neurons receive excitatory input from the ipsilateral ear and inhibitory input from the contralateral ear. From this input, different neurons in the LSO display a characteristic spiking rate when the sound sources are located at specific angles along the azimuthal plane. Finally, the output from the MSO and the LSO are integrated in the inferior colliculus (IC) [12], where neurons show a more coherent spatial representation across the entire audible frequency spectrum. The combination of both spatial cues can be seen as a multimodal integration process, where ITDs and ILDs are the modalities to be integrated in order to sharpen the neural representation of sound sources in the environment.

The importance of integrating ITDs and ILDs can be understood further by observing the topology of the IC, more specifically, by noting the overlap of MSO excitatory connections and LSO excitatory and inhibitory connections. On the one hand, the MSO can extract information about the sound source location from all sound frequencies, but it also produces noisy activity in higher frequencies. On the other hand, the LSO alone can extract information only from higher frequencies. For this reason, LSO excitatory connections to the IC reinforce informative activity from high frequencies in the MSO, while LSO inhibitory connections to the IC remove the noise produced by the MSO with high frequencies [14].

### B. Computational Background and Related Work

Large microphone arrays of different sizes and geometries are a common approach to SSL as they provide precise localization in multiple planes. These arrays can be designed to surround the space where the sound sources are located as in [15] and [16], or to be surrounded by the environment as it is the case of natural systems. The aim of our work is to explore the advantages of humanoid robotic platforms, hence, we focus on the latter case. The architecture proposed in [17] is an immersed array and can achieve a remarkable angular resolution of 3° with eight microphones. Similarly, the system described in [18] is designed with an array of 32 microphones and it is capable of localizing sound sources with an accuracy of 5° on the azimuth and elevation. The drawback of many approaches with large microphone arrays is that they only use the time difference of arrival (TDOA) between microphones for the estimation of sound sources. Since the information obtained from TDOAs is encoded most accurately in the low frequency components of sound, the performance of these systems depends on a small region of the audible sound spectrum. Furthermore, as these approaches use beamforming for speech segregation, the number of sound sources must be known in advance and the number of microphones has to be larger than the number of sound sources.

Acoustic daylight imaging [19] is an interesting approach that does not rely on TDOAs and can be used for SSL. However, similar to vision, this technique relies on the sound scattered by an object immersed in the noise field and is not capable of localizing the objects from directions where the array is not "looking" at. More recently, other SSL systems have been developed that can perform SSL robustly under a variety of noise and reverberation [20]–[22]. The architecture introduced in [22] is particularly interesting, as it has the ability to estimate the number of sound sources present in the environment. Part of their suggested future work includes an adaptive width for the window analyzing the input signals, as counting sound sources at a low signal-to-noise ratio (SNR) requires different parameters than at a high SNR. Yet, these systems also neglect the spatial information encoded in high frequencies of sound sources.

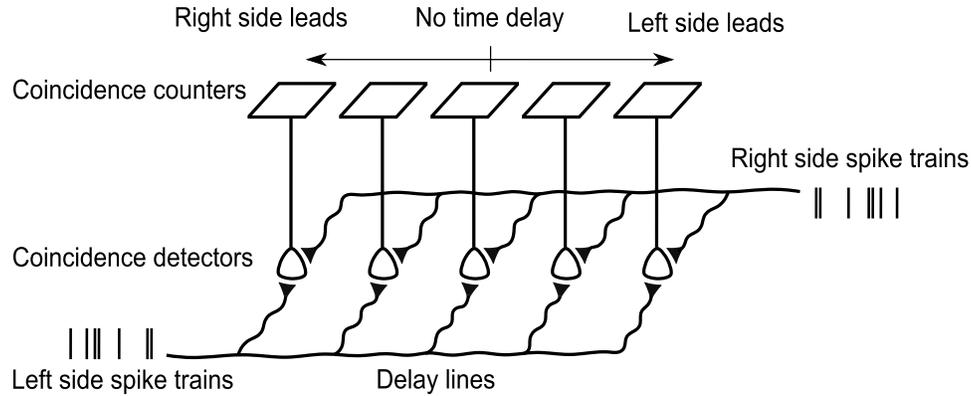

Fig. 2. Diagram of the MSO modeled as a Jeffress coincidence detector for representing ITDs [13]. This comparison is made between the spikes produced by the same frequency components $f$ when the time difference $\delta t$ between spikes is smaller than half a period, i.e., when $2f \cdot \delta t < 1$.

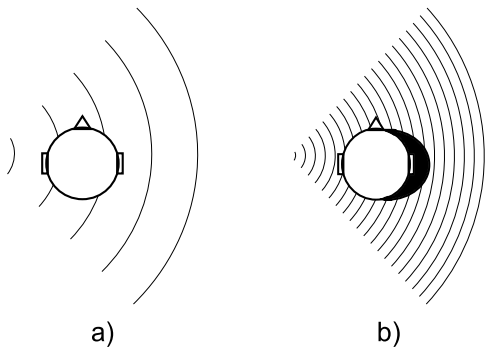

Fig. 3. (a) Interaction of a head structure and low frequency components in sound. (b) Interaction of a head structure and high frequency components in sound. Notice that a considerable shadowing effect is produced by the head only with high frequencies [4, Ch. 2.2.2].

An alternative to large microphone arrays is binaural SSL. With only one pair of microphones separated by a headlike structure, an SSL system can use ITDs and ILDs to locate sound sources in space. Both spatial cues are complementary, as ITDs convey more accurate information in low frequencies and ILDs in high frequencies. Fig. 3 shows the interaction between a headlike structure and different frequency components in sound. Integration of ITDs and ILDs is known as the *Duplex Theory of SSL*, and it places the boundary between low and high frequencies around 1500–3000 Hz [23]. The duplex theory can explain how the redundancy of information is achieved in natural SSL systems, as sounds in real-world environments are often rich in harmonic components. This redundancy can help to segregate information in noisy scenarios, such as outdoor environments or robotic platforms with intense ego noise [14].

The work introduced in [24] comes closer to the group of bioinspired binaural algorithms as the authors implement a multiple-delays model to estimate ITDs using artificial spiking neural networks (ASNN). Their system can localize broadband and low-frequency sounds with 30° accuracy, although its performance decreases for high-frequency sounds. An important advantage of ASNN is that they exploit the temporal dynamics in the sound signal, as the activation of a neuron depends on its current input and its previous activation state [25]. Furthermore, ASNN are biologically more plausible than other temporal neural models, and therefore, better suited for testing neurophysiological theories [26]. Rodemann *et al.* [27] developed a system that overcomes this limitation by including additional spatial cues. Their algorithm integrates ITDs, ILDs, and interaural envelope differences, and can localize the sound sources with a resolution of 10°, i.e., with three times finer granularity than the system in [24] using only one spatial cue. Nevertheless, the model in [27] shows high sensitivity to the ego noise produced by the robotic platform and requires further improvements to tackle this problem. Making use of neurophysiological principles from the mammalian auditory system, [28] and [29] describe probabilistic models of the MSO, LSO, and IC. Both systems show high SSL accuracy and can reach a resolution of 15°. A possible extension of this research is their implementation with ASNN in order to explore the dynamics of neural populations and to exploit their robustness against noise.

Liu *et al.* [30] model the MSO, LSO, and IC using ASNN, and the connection weights are calculated using Bayesian inference. Their system performs SSL with a resolution of 30° under reverberant conditions. In [14], we adapt the approach of [30] to the NAO robotic platform [31] with ∼40 dB of ego noise. This neural model is capable of handling such levels of ego noise and even increases the resolution of SSL to 15°. In more recent work, we compare several neural and statistical methods for the *representation*, *dimensionality-reduction*, *clustering*, and *classification* of auditory spatial cues [32]. The evaluation of these neural and statistical methods follows a tradeoff between computational performance, training time, and suitability for lifelong learning. However, the results of this comparison show that simpler architectures achieve the same accuracy as architectures with an additional clustering layer. Fig. 4 shows an overview of the best performing SSL architecture. We found that a neural classifier on the top layer of our architecture is important to increase the robustness of the system against the reverberation and ∼60 dB of ego noise produced by the humanoid iCub [33]. For this purpose, we include a feedforward neural network to handle the remaining nonlinearities in the output from the IC model. Finally, in order to improve the robustness of the system to data outliers, we extend our previous SSL system with softmax normalization on the output of the IC model and on the final layer of the SSL architecture.

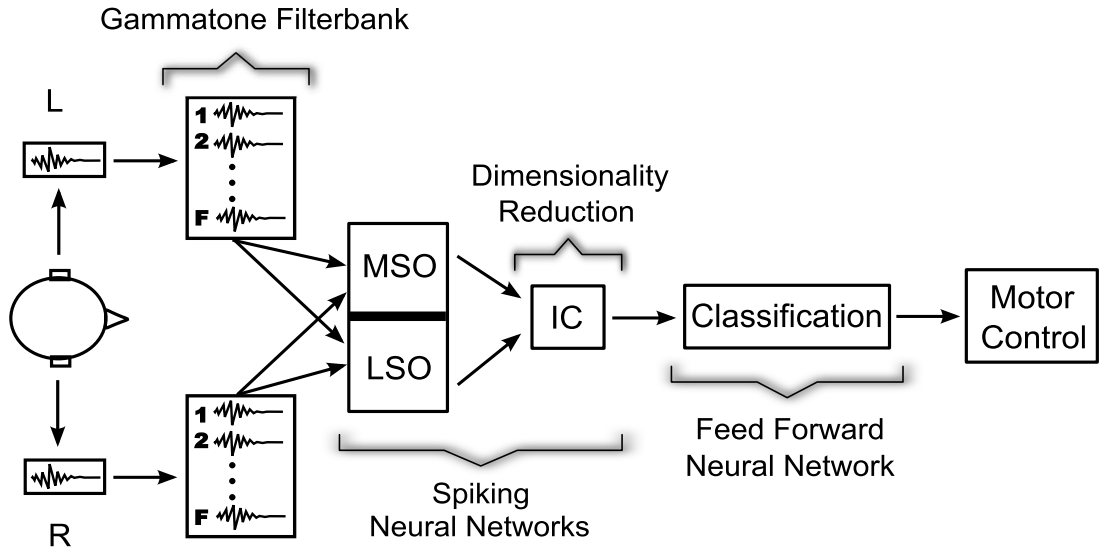

Fig. 4. SSL architecture. Sound preprocessing consists in decomposing the sound input in several frequency components with the Gammatone filterbank emulating the human cochlea [34]. Afterward, the MSO and LSO models *represent* ITDs and ILDs respectively. The IC model integrates the outputs from the MSO and LSO while performing *dimensionality reduction*. Finally, the *classification* layer produces an output angle that is used for motor control.

The following step in our research is to explore the use of SSL for improving the performance of ASR. Some interesting examples in this direction are presented in [35], [36], and [37]. These approaches make use of microphone arrays to localize the speech sources in the environment. Afterward, they use information about the sound source to separate the speech signals from noise in the background. The drawback of these methods is that they require prior knowledge about the presence and number of sound sources. [38] and [39] present two alternative approaches that make use of binaural robotic platforms. Yet, both systems suffer from the same limitations of the binaural SSL methods discussed before, as they mainly rely on information contained in low frequencies for SSL. Woodruff and Wang [40] present an interesting architecture, where they use ITDs and ILDs for SSL and can perform segregation of an unknown number of sources. Nevertheless, the reported results consider at most two sound sources, and segregation is performed offline due to the time required for computation. The approaches mentioned above rely on the construction of ideal binary masks for segregating speech. This presents an additional challenge because these methods are considerably affected when the sound source differs from the set of trained angles. Therefore, such approaches rely on an SSL system capable of tracking a human speaker almost instantly and with high accuracy. Our approach is focused on increasing the SNR of speech by continuously localizing the most intense sound source and reorienting the robot toward the speaker. In other words, we completely replace the use of ideal binary masks with a perception-action loop that maximizes the SNR of sound arriving from the direction of the speaker. Inspired by the paradigm of embodied cognition [41], [42], a key contribution from our work resides in shifting the focus of research toward maximizing the use of the humanoid embodiment: the robot can continuously increase the SNR of speech with the reflection from its pinnae to the microphone. This approach considerably reduces the computation by eliminating the use of binary masks and is feasible, given that our ASR system can recognize full sentences even if utterances have lower SNR at the beginning [43]. In order to compare more clearly the performance of ASR with and without the support of SSL, we constrain the domain-independent output of an ASR system to a domain-dependent set of sentences.

The paper is structured in the following way: in Section II, we describe in more detail each layer of our computational model for SSL and in Section III, we describe our experimental setup for testing SSL and ASR. More specifically, in Section III-A, we present the robotic platforms, in Section III-B, we introduce our virtual reality setup designed for experiments in cognitive robotics, and in SubSection III-C, we explain the mechanisms of our ASR system. In Section IV, we discuss the results of our experiments with static ASR and dynamic SSL and finally in Section V, we present our conclusions and future work.

## II. Bioinspired Computational Model

In this section, we briefly describe the SSL architecture based on our previous work in [30] and [14]. SSL is improved by applying a softmax normalization layer on the output of the IC model and a feedforward network for classifying the output of the IC model. Both are detailed at the end of this section. Further details on the virtual environment and the parameters of the architecture can be found in [44] and [32].

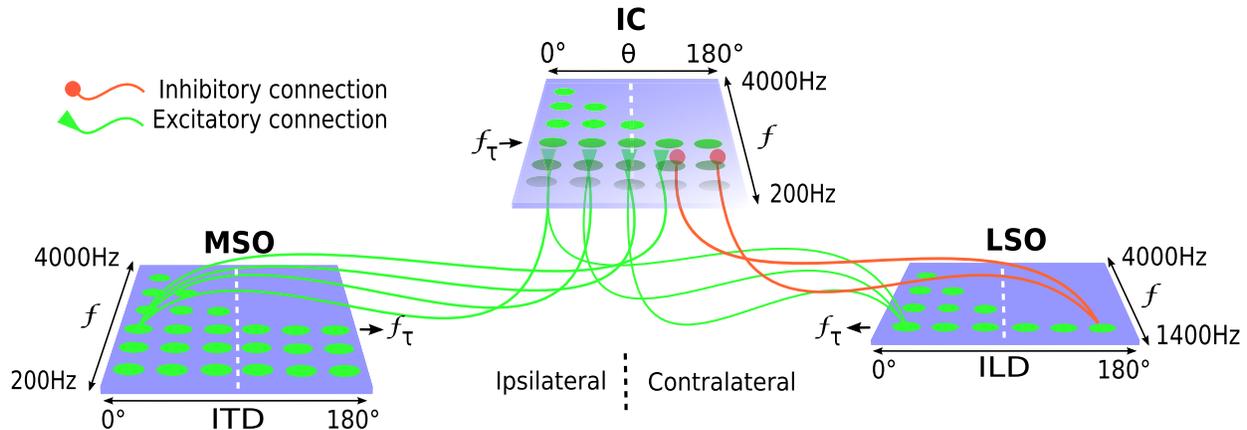

Fig. 5. Topology of the connections between the MSO and LSO models to the IC model. The MSO has excitatory connections to the IC in $f$ between 200 and 4000 Hz, whereas the LSO has excitatory and inhibitory connections to the IC only in $f \geq f_\tau$ between 1400 and 4000 Hz. Further details about the parameters used in the SNN model can be found in [30].

The first stage of our SSL architecture, shown in Fig. 4, consists of a gammatone filterbank modeling the frequency decomposition performed by the human cochlea [34]. This is, the signals produced by the microphones in the robot's ears are decomposed in a set of frequency components $f_i \in F = \{f_1, f_2, \ldots, f_I\}$. This tonotopic arrangement is preserved in all the subsequent layers in our SSL architecture. As we are mainly concerned with the localization of speech signals, we constrain the elements in $F$ to the frequency range where most speech harmonics are found, between 200 and 4000 Hz. Once both signals are decomposed into $I$ components (20 components as defined in [30]), each wave of frequency $f_i$ is used to generate spikes mimicking the phase-locking mechanism of the organ of Corti, i.e., a spike is produced when the positive side of the wave reaches its maximal amplitude.

In the following layer of the SSL architecture, we model MSO, where ITDs are represented. As depicted in Fig. 2, the computational principle observed in the MSO is modeled as a Jeffress coincidence detector [13] for each $f_i$. The MSO model has $m_j \in M = \{m_1, m_2, \ldots, m_J\}$ neurons for each $f_i$. The value of $m_J$ is constrained by the robot's interaural distance and the audio sampling rate. Each neuron $m_{i,j} \in \mathbb{N}^0$ is maximally sensitive to sounds produced at angle $\alpha_j$. Therefore, $\mathbf{S}^{\text{MSO}}$ is the array of spikes produced by the MSO model for a given sound window of length $\Delta T$. The mammalian auditory system relies mainly on delays smaller than half a period of each $f_i$ for the localization of sound sources [1, Ch. 5.3.3]. For this reason, the MSO model only computes ITDs when the time difference $\delta t$ between two incoming spikes is smaller than half a period, i.e., when $2 f_i \cdot \delta t < 1$. Inspired by the mammalian neuroanatomy, the MSO model projects excitatory input to all $f_i \in F$ of the IC model [45, Ch. 4, 6.].

At the same level of the SSL architecture, the LSO model represents ILDs. These are computed by comparing the L and R waves from each $f_i$ at the same points in time used for computing ITDs. Following the notation in Fig. 1, the $\log(A_1/A_2)$ of the amplitude values at times $t_1$ and $t_2$ determine the neuron in the LSO model that will fire. The LSO model has $l_j \in L = \{l_1, l_2, \ldots, l_J\}$ neurons for each $f_i$. As the value of $l_J$ is limited by the bit depth of the sound data, it is possible to have many more neurons in the LSO model than in the MSO model. For the sake of simplicity, we chose to have the same number of neurons in the MSO and LSO models by setting $l_J = m_J$. This decision does not have an impact on the system performance and establishes a clear boundary for the SSL granularity as the localization bins are the same for both spatial cues. Each neuron $l_{i,j} \in \mathbb{N}^0$ is maximally sensitive to sounds produced at angle $\alpha_j$. Therefore, $\mathbf{S}^{\text{LSO}}$ is the array of spikes produced by the MSO model for a given sound window of length $\Delta T$. Also inspired by the mammalian neuroanatomy, the LSO model projects excitatory and inhibitory input only to the highest frequencies of the IC model $f_i \in F \mid f_i \geq f_\tau$; where the threshold $f_\tau = 1400$ Hz [45, Ch. 4, 6.].

Then, we arrive at the layer modeling the IC, where ITDs and ILDs are integrated. The topology of the connections between the MSO and LSO models to the IC model can be seen in Fig. 5. Bayesian classifiers allow the continuous update of probability estimations and are known to have good performance even under strong independence assumptions. Furthermore, Bayesian classifiers allow fast computation as they can extract information from large dimensional data in a single batch step. For this reason, we estimate the connection weights assigned to the excitatory and inhibitory output of the MSO and LSO layer using Bayesian inference [30]. The IC model has $c_k \in C = \{c_1, c_2, \ldots, c_K\}$ neurons for each $f_i$. Each neuron $c_{i,k} \in \mathbb{R}$ is maximally sensitive to sounds produced at angle $\theta_k \in \Theta_K = \{\theta_1, \theta_2, \ldots, \theta_K\}$, where $K$ is the total number of angles around the robot where sounds were presented for training. $\mathbf{E}^{\text{MSO}}$ and $\mathbf{E}^{\text{LSO}}$ are the ipsilateral MSO and LSO excitatory connection weights to the IC, and $\mathbf{I}^{\text{LSO}}$ are the contralateral LSO inhibitory connection weights to the IC. Therefore, $\mathbf{S}^{IC}$ is the array of spikes produced by the IC model for a given sound window of length $\Delta T$. More precisely, $\mathbf{S}^{IC}$ is computed by adding the elementwise product of the following matrices:

$$\mathbf{S}^{IC} = \mathbf{S}^{MSO} \odot \mathbf{E}^{MSO} + \mathbf{S}^{LSO} \odot \mathbf{E}^{LSO} - \mathbf{S}^{LSO} \odot \mathbf{I}^{LSO}. \quad (1)$$

In order to estimate the connection weights $\mathbf{E}^{MSO}$, $\mathbf{E}^{LSO}$, and $\mathbf{I}^{LSO}$, we perform Bayesian inference on the spiking activity $\mathbf{S}^{MSO}$ and $\mathbf{S}^{LSO}$ for the known sound source angles $\Theta_K$.

We define the set of training matrices obtained for each $\theta_k$ as $s_n \in S = \{s_1, s_2, \ldots, s_N\}$, where $N$ is the total number of training instances. We describe first the Bayesian process used to estimate the connection weights between the MSO and the IC, where $s_n = \mathbf{S}_n^{MSO}$. Let $p(\mathbf{S}^{MSO}|\theta_k)$ be the likelihood that a sound that occurs at angle $\theta_k$ produces the spiking matrix $\mathbf{S}^{MSO}$. As we assume Poisson distributed noise in the activity of neurons $m_{i,j}$ in the MSO model

$$p(\mathbf{S}^{MSO}|\theta_k) = \frac{\lambda_k \exp^{-\lambda_k}}{\mathbf{S}^{MSO}!}, \quad \forall k \in \Theta_K, \quad (2)$$

where $\lambda_k$ is a matrix containing the expected value and variance of each neuron $m_{ij}$ in $\mathbf{S}^{MSO}$, and it is computed from the training set $S$ for each $\theta_k$. In a Poisson distribution, the maximum likelihood estimation of $\lambda_k$ is equal to the sample mean and is calculated as

$$\lambda_k = \frac{1}{N} \sum_{n=1}^{N} \mathbf{S}_n^{MSO}, \quad \forall s_n \in S \mid \theta_k. \quad (3)$$

As we assume a uniform distribution over all angles in $\Theta_K$, we assign the same prior $p(\theta_k) = 1/K$ to each $\theta_k$. In order to normalize the probabilities to the interval $[0, 1]$, we compute the evidence $p(\mathbf{S}^{MSO})$ as

$$p(\mathbf{S}^{MSO}) = \sum_{k=1}^{K} p(\mathbf{S}^{MSO}|\theta_k) p(\theta_k). \quad (4)$$

Afterward, the posterior $p(\theta_k|\mathbf{S}^{MSO})$ is computed using Bayes rule

$$p(\theta_k|\mathbf{S}^{MSO}) = \frac{p(\mathbf{S}^{MSO}|\theta_k) p(\theta_k)}{p(\mathbf{S}^{MSO})} = \mathbf{P}_k^{MSO}. \quad (5)$$

The same Bayesian inference process described so far is used for computing the LSO inhibitory and excitatory connections to the IC. Finally, the connection weights for each neuron $m_{i,j}$ in $\mathbf{P}_k^{MSO}$ and $l_{i,j}$ in $\mathbf{P}_k^{LSO}$ to neuron $c_{i,k}$ in the IC are set according to the following functions:

$$\mathbf{E}^{MSO} = \begin{cases} \mathbf{P}_k^{MSO}, & \text{if } \mathbf{P}_k^{MSO} > \\ & (\omega_E^{MSO} \cdot \arg\max_{\theta_k}(\mathbf{P}_k^{MSO})) \\ 0 & \text{otherwise} \end{cases} \quad (6)$$

$$\mathbf{E}^{LSO} = \begin{cases} \mathbf{P}_k^{LSO}, & \text{if } \mathbf{P}_k^{LSO} > \\ & (\omega_E^{LSO} \cdot \arg\max_{\theta_k}(\mathbf{P}_k^{LSO})) \\ & \bigwedge f_i \geq f_\tau \\ 0 & \text{otherwise} \end{cases} \quad (7)$$

$$\mathbf{I}^{LSO} = \begin{cases} 1 - \mathbf{P}_k^{LSO}, & \text{if } \mathbf{P}_k^{LSO} < \\ & (\omega_I^{LSO} \cdot \arg\max_{\theta_k}(\mathbf{P}_k^{LSO})) \\ & \bigwedge f_i \geq f_\tau \\ 0 & \text{otherwise,} \end{cases} \quad (8)$$

where $\omega_E^{MSO} \wedge \omega_E^{LSO} \wedge \omega_I^{LSO} : \mathbb{R} \in [0, 1]$ are scalar thresholds that determine which connections will be pruned. In accordance to known neuroanatomy, such pruning avoids interaction between THE neurons sensitive to distant angles [46]. The value of $f_\tau$ marks the transition between the lower and higher frequency spectra.

Finally, we use a feedforward neural network in the last layer of our SSL system for the classification of $\mathbf{S}^{IC}$. This layer is important for providing the system robustness against ego noise and reverberation. The output of the IC layer still shows nonlinearities that reflect the complex interaction between the robot's embodiment and sound in the environment. Some of the elements that influence this interaction include the sound source angle relative to the robot's face, the head material and geometry, and intense levels of noise produced by the cooling system inside the robot's head. In previous work, we compare several neural and statistical methods [32] and found that a multilayer perceptron (MLP) was the most robust method for representing the nonlinearities in $\mathbf{S}^{IC}$. The hidden layer of the MLP performs compression of its input as it has $|\mathbf{S}^{IC}|/2$ neurons, and similar to the IC neurons analyzing a single $f_i$, the output layer of the MLP has $c_k \in C$ neurons. In order to improve the robustness of the system against data outliers, we perform softmax normalization on $\mathbf{S}^{IC}$ before training the MLP

$$\mathcal{S}^{IC} = \left( \frac{\exp^{\mathbf{S}_i^{IC}}}{\sum_{i'=1}^{I'} \exp^{\mathbf{S}_{i'}^{IC}}} \right), \quad \forall f_i \in F, \quad (9)$$

and also on the output $\mathbf{S}^{MLP}$ of the MLP

$$\mathcal{S}^{MLP} = \max_k \left( \frac{\exp^{\mathbf{S}_k^{MLP}}}{\sum_{k'=1}^{K'} \exp^{\mathbf{S}_{k'}^{MLP}}} \right), \quad \forall c_k \in C. \quad (10)$$

Fig. 6 shows the output of all layers in our SSL architecture after training it with a subset of utterances from the Texas Instruments—Massachusetts Institute of Technology (TIMIT) speech data set [47]. The figures show the spiking matrices produced with white noise in order to depict more clearly the stereotypical patterns of each $f_i$. Notice that the hypotheses generated by most neurons in the IC layer agree on the sound source angle, irrespective of the frequency component $f_i$ they receive input from. In this case, it is not surprising that the MLP classifies correctly $\mathbf{S}^{IC}$, since using the *winner-takes-all* rule along each $f_i$ would suffice for correct classification. Further details about the parameters of the SSL architecture and the training methodology can be found in [30] and [32].

## III. EXPERIMENTAL SETUP AND BASIS METHODOLOGIES

### A. Humanoid Robotic Platforms

In our experiments, we use two different humanoid robotic heads: *iCub* [33] and *Soundman* [48]. A lateral view of both platforms and their pinnae can be seen in Fig. 7. The iCub is a humanoid robot designed for research in cognitive developmental robotics. Its head is made of a plastic skull and contains electronic and mechanical components, including a fan that continuously produces ~60 dB of ego noise. Soundman is a commercial dummy head designed for the production of binaural recordings that increase the perception of spatial effects. This head is made of solid wood, has no interior components, and hence, does not produce ego noise. We added a motor to the head that allows it to rotate on the yaw axis. Sound spatial cues are produced by the geometric and material properties of the humanoid heads, and both platforms allow the extraction of sound spatial cues from binaural recordings. The objective of using both heads is to measure the performance of SSL and ASR with Soundman, and use these measurements as a performance baseline for the iCub. This comparison allows to determine if the resonance from the skull and components inside the iCub head reduce the performance of SSL and ASR.

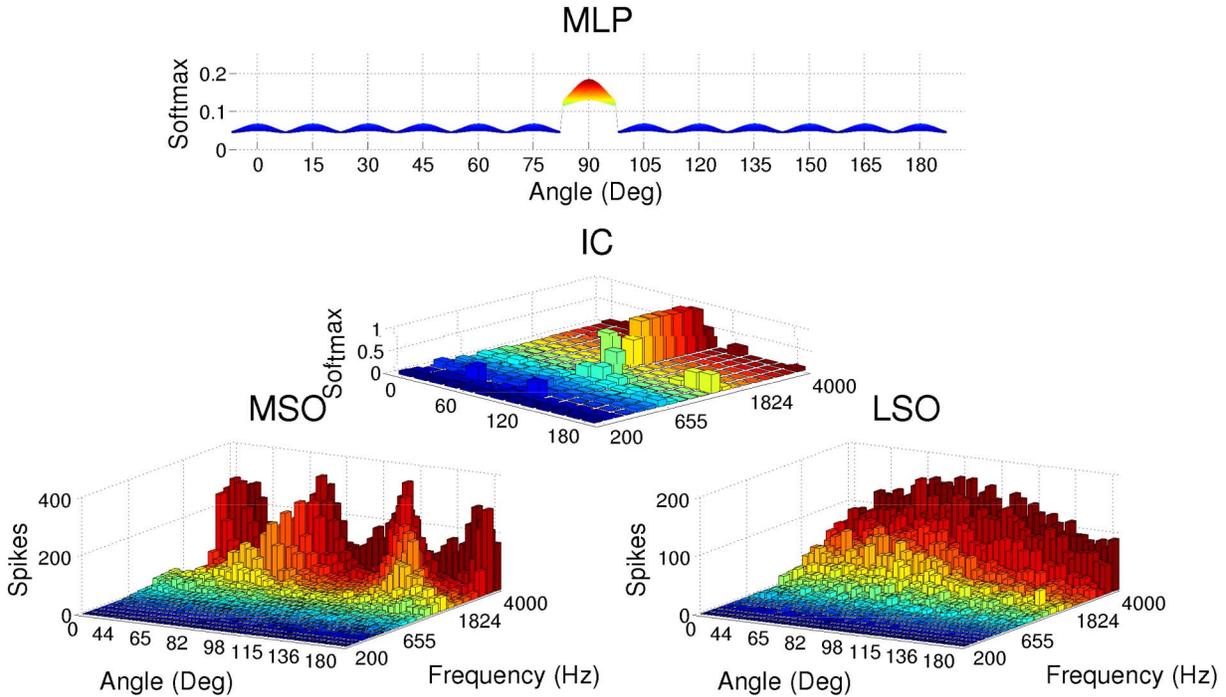

Fig. 6. Output of all the layers in the SSL architecture for white noise presented in front of the robot (90°). Notice that for this angle, most of the IC frequency components agree on the sound source angle, and the MLP correctly classifies the IC output.

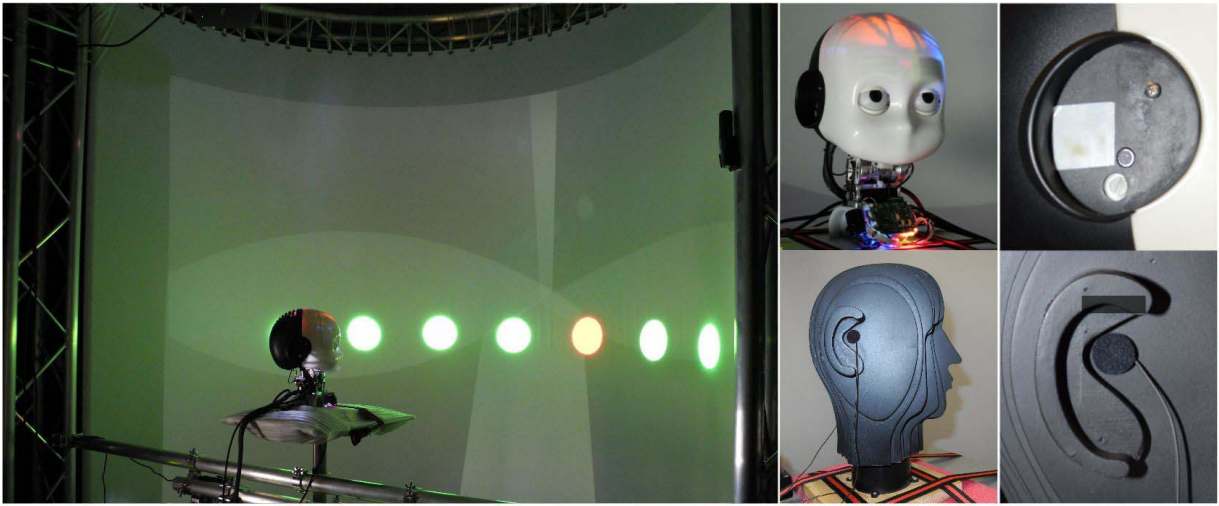

Fig. 7. Left: audio-visual virtual reality experimental setup. The light blobs show the curvature of the *half-cylinder* projection screen surrounding the iCub humanoid head and represent the location of sound sources behind the screen. Right: both humanoid robotic heads used during our experiments and a zoom to their ears. The robots' ears consist of microphones perpendicular to the sagittal plane and are surrounded by pinnae. Further details about the VR setup and the principles that guide its design can be found in [44].

### B. Virtual Reality Setup

We perform the experiments in an audio-visual virtual reality (VR) setup designed by our group for the development of multimodal integration systems. In the VR setup, it is possible to control the temporal and spatial presentation of images and sounds to different robotic platforms. As we see in Fig. 7, the humanoid is located at the radial center of a projection screen shaped as a half cylinder and the noise produced by the projectors is below 30 dB at the location of the robot. The auditory stimuli used for the experiments described in this paper are described in Section III-C. These auditory stimuli are presented from 13 loudspeakers evenly distributed on the same azimuth plane at angles $\theta_{\text{lspk}} \in \{0°, 15°, \ldots, 180°\}$ and the loudspeakers are placed behind the screen at $\sim$1.6 m from the robot. The room acoustics are partially damped by corrugated curtains in order to approach a reverberation time (0.25–0.5 s) and an inner sound pressure level (20–40 dB) with *studio* quality. When we perform ASR experiments with iCub OFF or when we use Soundman, the same pair of balanced microphones is mounted on either head and the sound stimuli have an intensity of $\sim$60 dB. When we perform SSL experiments with iCub ON, the intensity of the sound stimuli are increased to $\sim$80 dB due to the high levels of ego noise produced by the robot. Further details about the VR setup and the principles that guide its design can be found in [44].

### C. Automatic Speech Recognition System

We use a system developed by our group for ASR [43]: Domain- and Cloud-based Knowledge for Speech Recognition (DOCKS). The DOCKS system has two main components: 1) A *domain-independent* speech recognition module and 2) a *domain-dependent* phonetic postprocessing module. The need for domain-dependent ASR arises from the intense noise of the cooling system in humanoid platforms commonly used for research in academia (NAO, iCub). In such conditions, sentences are more easily recognizable than words, which is analogous to the British Royal Air Force alphabet used in aviation to communicate under low SNR conditions. The domain-dependent output of the DOCKS system does not impede generalisation from our experimental results, as our objective is not to develop a novel ASR system. Our goal is to compare the performance of any existing ASR system *with* and *without* the support of SSL.

To test the DOCKS ASR system, Heinrich and Wermter [49] created a corpus that contains 592 utterances produced from a predefined grammar. The corpus was recorded by female and male nonnative speakers using headset microphones, and it is especially useful as the grammar for parsing the utterances is available. Two commercial ASR platforms were chosen as the domain-independent component of the DOCKS system: Google ASR [50] and Sphinx [51]. Both are compared by measuring the word error rate (WER) and sentence error rate (SER) under four different configurations. In Table I, we compare the performance of: 1) the raw output of *Google ASR* (Go); 2) *Sphinx ASR* (Sp) with an N-Gram (NG) language model, with the corpus finite state grammar (FSG) and with the domain sentences (DoSe); 3) Go plus the *Sphinx Hidden Markov Model* (Sp-HMM) with NG, with FSG and with DoSe; and 4) Go with the domain word list (WoLi) and with the domain sentence list (SeLi).

During the *domain-independent* speech recognition, the DOCKS system uses *Go*. As in previous work [52], it has shown better performance than *Sp*. In our experiments, we use the TIMIT core-test-set (TIMIT-CTS) [47] as speech stimuli. The TIMIT-CTS is formed by the smallest TIMIT subset that contains all the existing phonemes in the English language. It consists of 192 sentences spoken by 24 different speakers: 16 male and 8 female pronouncing 8 sentences each. Further details about the DOCKS architecture can be found in [43] and [32].

TABLE I
PERFORMANCE OF ASR SYSTEMS

| System | WER in % | SER in % |
|---|---|---|
| Go | 50.230 | 97.804 |
| Sp + NG | 60.462 | 95.101 |
| Sp + FSG | 65.346 | 85.980 |
| Sp + DoSe | 65.346 | 85.980 |
| Go + Sp-HMM + NG | 7.962 | 27.703 |
| Go + Sp-HMM + FSG | 6.038 | 19.257 |
| Go + Sp-HMM + DoSe | 5.846 | 18.581 |
| Go + WoLi | 23.231 | 57.432 |
| Go + SeLi (**DOCKS**) | **3.077** | **11.993** |

Best results are marked in boldface.
Terminology can be found in the text.

During the *domain-dependent* phonetic postprocessing, the DOCKS system maps the output of *Go* to the sentences in the TIMIT-CTS. Whenever a sound file is sent to *Go*, a list with the 10 most plausible sentences (G10) is returned. First, the system transforms the G10 and the TIMIT-CTS from *grapheme* representation to *phoneme* representation [53]. Then, the system computes the Levenshtein distance [54] between each of the phoneme sequences in the G10 and the TIMIT-CTS. Finally, the phoneme sequence in the TIMIT-CTS with the smallest distance to any of the phoneme sequences in the G10 is considered the winning result. The sentence corresponding to the winning phoneme sequence is considered correct when it matches the ground truth sentence presented to the robot.

### IV. EXPERIMENTAL RESULTS AND DISCUSSION

### A. Optimal Sound Source Direction for Speech Recognition

The objective of this experiment is to compare the effect of shadowing from both humanoid heads on the SNR of speech stimuli and to find the optimal facing angle for ASR. In addition to our architecture proposed in [32], we added a softmax normalization to the output of the IC model and to the feedforward network in the last layer of the architecture. These extensions increase the robustness of the system against outliers. Let $\theta_{\text{neck}}$ be the angle faced by the robot at any given time, $\theta_{\text{lspk}}$ the fixed angle of the loudspeakers producing the stimuli, and $\delta_{\text{diff}}$ is the angular distance between $\theta_{\text{lspk}}$ and $\theta_{\text{neck}}$. We hypothesise that there is a subset of angular distances $\delta_{\text{best}} \subset \delta_{\text{diff}}$ for which the SNR of sensed speech is highest, and hence, for which the DOCKS system performs the best when using the humanoid heads. In order to find $\delta_{\text{best}}$, we present 10 times the entire TIMIT-CTS corpus around the humanoid heads from each of the loudspeakers at angles $\theta_{\text{lspk}}$ while keeping $\theta_{\text{neck}}$ fixed. Then, we measure the DOCKS system performance as the average SER of speech recognition for each $\delta_{\text{diff}}$. We define SER as the ratio of incorrect recognitions (*false positives*) over the total number of recognitions (*true positives + false positives*). It is also interesting to compare this *binary* measure with a *continuous* measure of performance. We can make such comparison by observing the Levenshtein distance between the output of the DOCKS system and the ground truth sentences.

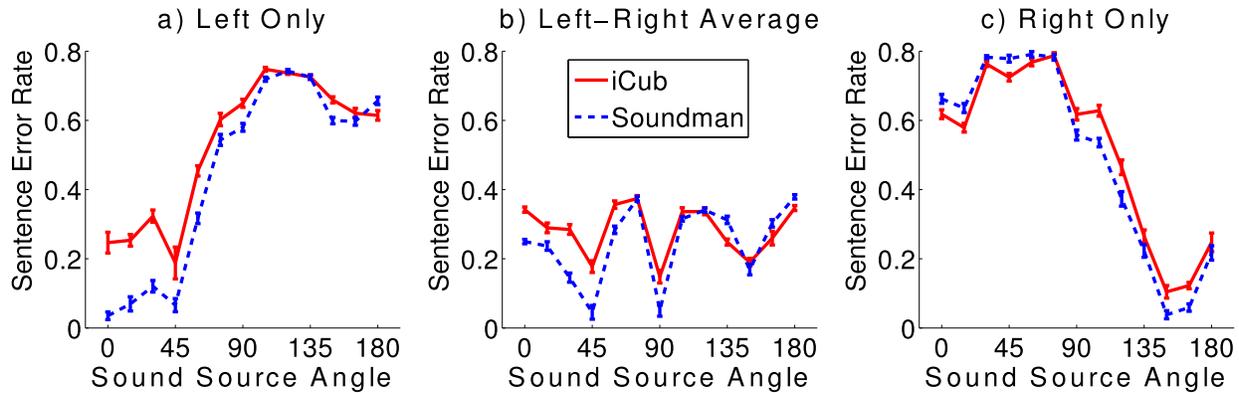

Fig. 8. **Binary measure of ASR performance.** Average SERs of the DOCKS system for recognizing utterances presented at various angles. The legend in the middle applies to the three figures and the bars at each point represent the standard deviation over the ten trials. The results were obtained with both robotic heads for the frontal 180° on the azimuth plane.

As most ASR engines, the DOCKS system requires monaural files as input. Therefore, the stereo recordings made with the robotic heads are *reduced* to one channel. There are three possible downmixing procedures: 1) using the sound wave from the left channel only (LCh); 2) using the sound wave from the right channel only (RCh); or 3) averaging the sound waves from both channels (LRCh). Fig. 8 shows the average SERs of the DOCKS system with the three downmixing procedures using both humanoid heads. The bars at each point represent the standard deviation over the 10 trials. Similarly, Fig. 9 shows the average Levenshtein distances between the output of the DOCKS system and the ground truth sentences. These are the distances that were used to produce the binary results shown in Fig. 8, which explains the resemblance of their shape and confirms the close relation between SERs and distances in the Levenshtein space.

The *smoothness* and symmetry of the curves is possibly affected by several factors including: varying reverberation, different fidelity of each loudspeaker, asymmetry between the left and right pinnae of the iCub and imbalances between the left and right microphones. Nevertheless, the results obtained with the three downmixing procedures corroborate the existence of similar $\delta_{\text{best}}$ for both robotic heads. More specifically, the DOCKS system has a considerably better performance at $\delta_{\text{best}} \in \{\sim 45°, \sim 150°\}$. The performance of speech recognition is affected by the SNR of speech, and the SNR of speech is affected by the directional shadowing produced by the head. Therefore, as the performance curves of the DOCKS system are very similar with the recordings from both heads, we conclude that the structural, geometrical, and material properties of the iCub head produce a directional shadowing very similar to the one produced by Soundman. These results confirm the effectiveness of the iCub for the production of spatial cues.

Before running the experiment, we expected the speech SNR to be maximal when the sound source is parallel to the interaural axis, i.e., for $\theta_{\text{lspk}} \in \{0°, 180°\}$. Surprisingly, both angles $\delta_{\text{best}}$ are located $\sim 45°$ to the left and right of the sagittal plane. This effect could be produced by the reflection of sound waves from the pinna toward the microphone closest to the sound source. In this case, $\delta_{\text{best}}$ could be the angles where such reflection is most intense. Due to the head shadowing, recordings only have the same SNR on both channels when the sound source is placed exactly in front of the robot. In all other angles $\delta_{\text{diff}}$, the microphone closest to the sound source records with higher SNR than the other one. For this reason, the LRCh downmixing diminishes the SNR of speech after both signals are averaged. Together, the head shadowing and the pinnae reflection explain why the DOCKS performs best at 45°, 90° and 150° in the LRCh downmixing.

It is also important to note that the lowest SERs from the LCh and RCh downmixings are about *twice* as large as the lowest SERs from the LRCh downmixing. This substantial increase in performance is possible because in the LCh and RCh downmixings, the channel with higher SNR remains uncorrupted by the signal from the channel with lower SNR. It is interesting to note that all figures of the LCh and RCh downmixings show a periodical shape. This phenomenon could be caused by the circular shape of the humanoid heads and the position of the microphones. As both pinnae are placed slightly behind the midcoronal plane, the distance traveled by sound waves from the sound source to the furthest ear is maximal at ~45° or at ~150°. This configuration explains the slight SER decrease after 135° with LCh and before 30° with RCh.

### B. Dynamic Sound Source Localization

When we say that SSL can help to improve the performance of the DOCKS system, we assume that the robot will turn to the optimal listening angle in a small number of localization steps or *SSL iterations*. Furthermore, once the robot is optimally oriented it should remain stable in such position, or proceed to track the speech source closely as soon as it moves around it. The objective of this experiment is to find how many SSL iterations it takes the system to face a sound source, starting from different angles between the sound source and the direction faced by the robot. Once the robot is facing directly at the sound source, we can measure the stability of the SSL system for *locking* on the speech target. It is important to measure this locking on each of the 13 loudspeakers in the VR setup at angles $\theta_{lspk}$ in order to verify that the SSL system is robust to the reverberation produced in different room locations around the robot. During the experiment, we present the robot with a sound composed of utterances from 24 different speakers: 16 males and 8 females. More specifically, the longest sentence from each speaker in the TIMIT-CTS corpus is appended in a single sequence of utterances to form a 106 s *compound sound*. Once a compound sound is formed, the last two sentences of the sequence of utterances are moved to the beginning, creating another compound sound. By repeating the same procedure, 12 compound sounds are produced in total.

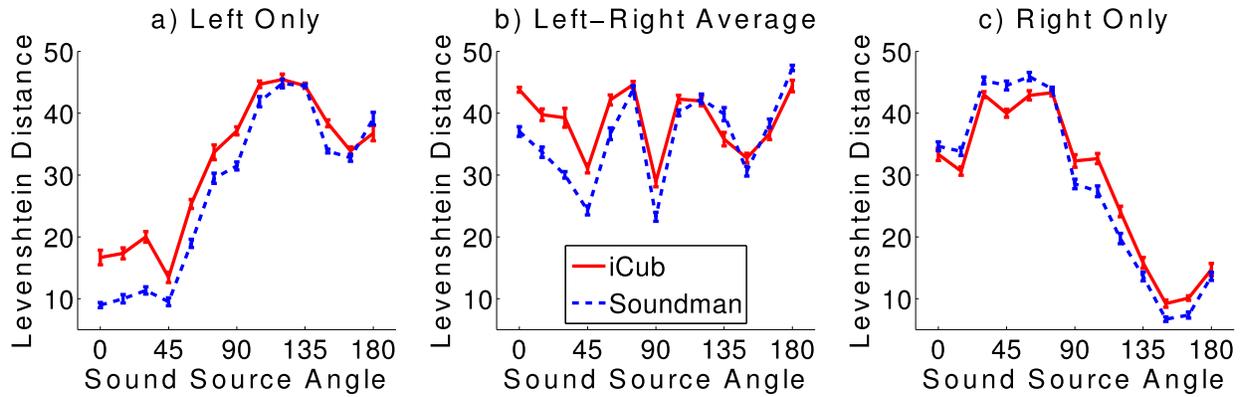

Fig. 9. **Continuous measure of ASR performance.** Average Levenshtein distances between the DOCKS output and the ground truth for sentences presented at various angles. The legend in the middle applies to the three figures and the bars at each point represent the standard deviation over the 10 trials. The results were obtained with both robotic heads for the frontal 180° on the azimuth plane. Notice that the edit distance allows us to see that, even in the best cases, the Levenshtein distance is greater than zero, i.e., none of the sentences would be recognized without the domain-dependent component of our ASR system. Reprinted by permission from Springer Nature: Springer Lecture Notes in Computer Science, *J. Dávila-Chacón, J. Liu and S. Wermter, Improving Humanoid Robot Speech Recognition with Sound Source Localisation*, © Springer International Publishing Switzerland 2014.

At the beginning of each trial, the robot turns to a starting neck angle $\theta_{neck} \in \{45°, 15°, \ldots, 135°\}$ on the azimuth plane. The starting angles $\theta_{neck}$ are constrained by the turning limitations of the yaw joint in the robot's neck. Once the robot is oriented in the first $\theta_{neck}$, the first compound sound is reproduced from the loudspeaker at angle $\theta_{lspk}$ and the robot starts tracking the sound source. The trial ends when the sound finishes. Then, the robot head returns to the same angle $\theta_{neck}$ and the same compound sound is now presented at the following loudspeaker. This procedure is repeated until all angles $\theta_{lspk}$ are covered. Afterward, the same routine over all angles $\theta_{lspk}$ is repeated for each starting angle $\theta_{neck}$. Finally, the entire process is repeated for each of the 12 compound sounds. This procedure is necessary in order to discard the possibility that the voice of a particular speaker systematically affects the SSL system at the same point in time.

The results of the dynamic localization task are summarized in Fig. 10(a) for iCub and in Fig. 11(a) for Soundman. The figures show the performance of the SSL system in consecutive iterations and from a range of starting angular differences between $\theta_{neck}$ and $\theta_{lspk}$, where $\delta_{start} \in \{0°, 15°, \ldots, 90°\}$. The dotted lines in both figures show the average SSL performance of trials with the same starting angular difference $\delta_{start}$. The continuous lines show the average and standard deviations of all starting angular differences $\delta_{start}$. In both figures, it can be seen that the localization error decreases as $\delta_{start}$ decreases from 90° to 0°. The curves show that the system converges to the sound source angle in 3 iterations or less. Afterward, localization errors are close to zero with almost no variance. In other words, the SSL system is more robust for localizing sounds closer to the front of the head. As localization errors are smaller in the frontal angles, the SSL system converges to the sound source angle after successive localization steps. Once the robot is facing the sound source, it continues facing that direction, i.e., the SSL system successfully locks the auditory target. These results are consistent with our previous work on static SSL [14], [32] and with the performance observed in humans [23].

Figs. 10(b) and 11(b) show the angular error accumulated from all SSL iterations. During the experiments, many more data points were produced for angles $\delta_{diff}$ close to 0°. However, the variance of the accumulated errors also indicates better SSL performance when the sound source is close to the frontal angles. Importantly, this improvement applies to all angles $\theta_{lspk}$. This consistency in performance shows the robustness of our architecture against the changes in reverberation produced by presenting auditory stimuli from different room locations. Therefore, we conclude that the proposed SSL architecture successfully avoids overfitting to the training data from static sound sources and does not stagnate in poor local minima. It is also important to note that the magnitude of localization errors is related to the size of the chosen localization bins (15° of angular granularity). Nevertheless, some preliminary studies show that our system is capable of 1° angular resolution in the frontal 40°. We could access this potential by performing SSL in a continuous space using the last layer for regression instead of classification. Verifying this hypothesis is part of our following work with the SSL architecture. Finally, we conclude that the difference in performance between both robotic heads reflects the additional challenges present in the iCub due to the intense ego noise. Nevertheless, the system reaches near-perfect accuracy once the sound source is located within 30° from the frontal angle with both platforms.

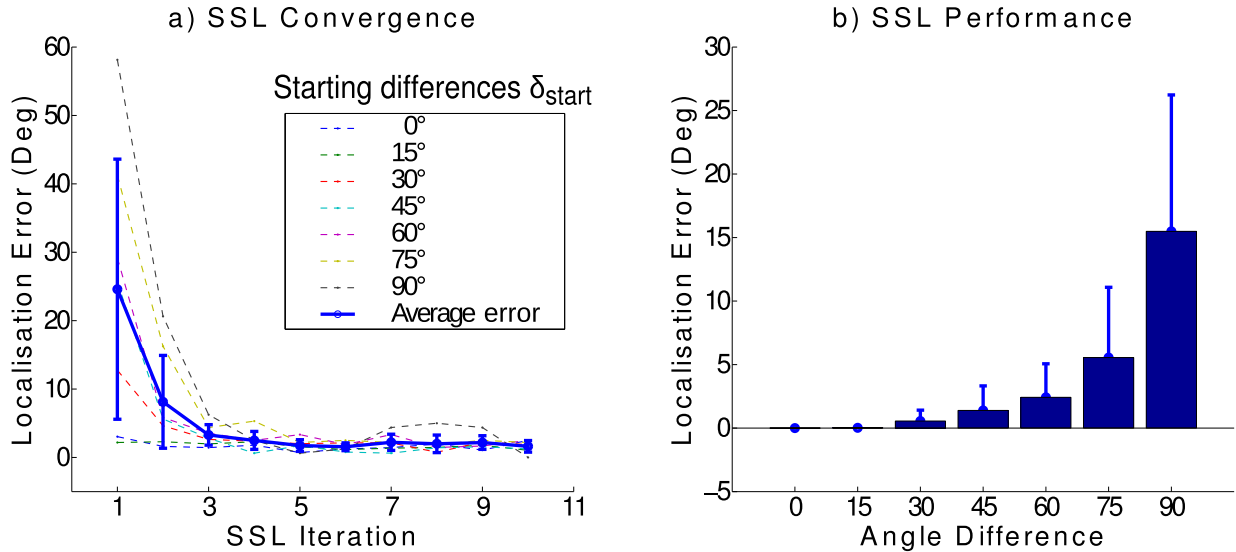

Fig. 10. **Dynamic SSL using the iCub head.** (a) SSL performance in consecutive iterations. The dotted curves display the performance for a range of starting angular differences. At each trial, a composed speech recording is presented to the robot. The solid line shows the average of all dotted curves with the bars indicating the standard deviation. Note the small number of steps required for the robot to reach near 0 error, i.e., to face the correct sound source angle. (b) Accumulated angular error from all iterations in all SSL trials. Note that the accuracy of the SSL system is higher when the angle difference between the sound source and the direction faced by the robot is 0, i.e., when the robot is facing the sound source.

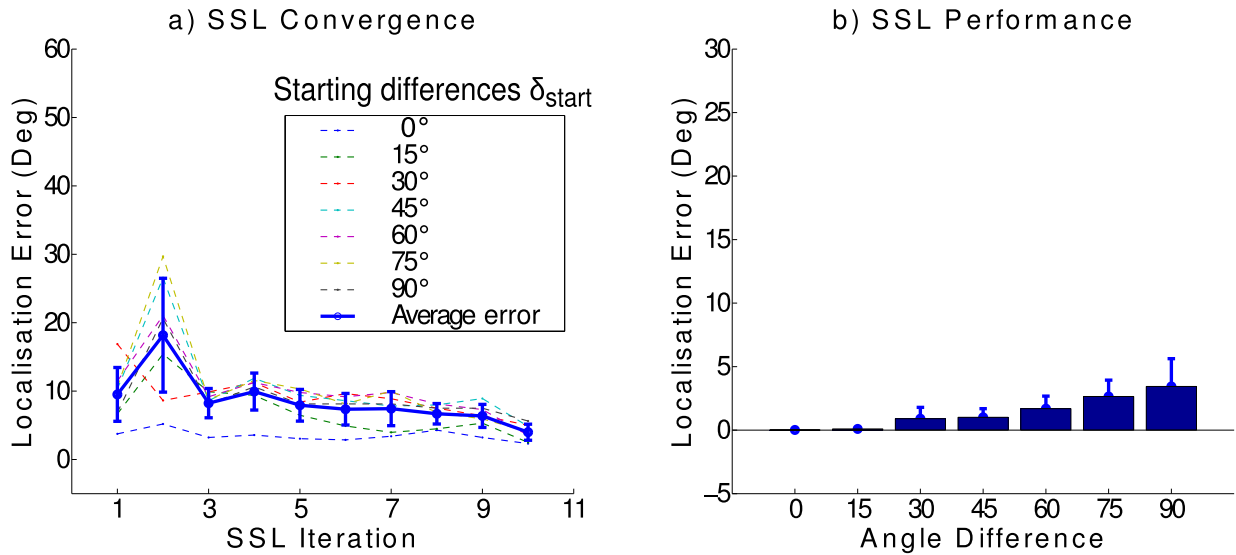

Fig. 11. **Dynamic SSL using the Soundman wooden head.** (a) SSL performance in consecutive iterations. The dotted lines display the performance for different angular differences at the beginning of each trial presenting a composed speech sound to the robot. The solid line shows the average of all dotted curves with the bars indicating the standard deviation. (b) Accumulated angular error from all iterations in all SSL trials.

## V. CONCLUSION AND FUTURE WORK

From the experimental results, we found that using information from SSL can improve considerably the accuracy of speech recognition for humanoid robots. As the humanoid platform provides signals from the left and right channels, SSL can indicate how to orient the robot, and then, select the appropriate channel as input to an ASR system. This approach is in contrast to related approaches where signals from both channels are averaged before being used for ASR. Our proposed method is capable of doubling the recognition rates at the sentence level when compared to the common averaging method. Interestingly, the performance of the ASR system is not highest when the sound source is facing directly to the microphone in one of the humanoid's ears, but at the angle where the pinna reflects most intensely the sound waves to the microphone. It is possible to measure the magnitude of this improvement by repeating the ASR experiment with the pinnae removed from the heads.

The results of the dynamic SSL experiment show that the architecture is capable of handling different kinds of reverberation. These results are an important extension from our previous work in static SSL and support the robustness of the system to the sound dynamics in real-world environments. Furthermore, our system can be easily integrated with recent methods to enhance ASR in reverberant environments [55]–[57] without adding computational cost. This is the intrinsic advantage of embodied embedded cognition. As another extension considering the dynamics of real-world scenarios, we plan to embed the SSL architecture into a probabilistic framework. In this framework, time will be integrated in the estimation of sound source angles by using calculations from previous time steps to increase the confidence of the system estimations. This probabilistic model will also benefit from a parallelised version of the MSO and LSO spiking neural layers. In a preliminary GPU implementation, we have already reached 12 times more SSL iterations in the same amount of time than the current CPU version.

An important advantage of our biomimetic neural representation of spatial cues is that it can be directly integrated with vision for audio-visual spatial attention [58]. In this scenario, vision can be used to disambiguate the location of a sound source of interest in a cluttered auditory landscape. As each frequency component generates a spatial hypothesis in our IC model, vision can be used to perform auditory grouping in the time and frequency domains [59], [60]. Furthermore, vision can also be used as a bootstrapping mechanism for training the neural layers in an online fashion. In this way, the entire architecture can be trained with an unsupervised learning approach. This is the main direction of our current research toward multimodal speech recognition.

## References

[1] J. Schnupp, I. Nelken, and A. J. King, *Auditory Neuroscience: Making Sense of Sound*. Cambridge, MA, USA: MIT Press, 2012.
[2] M. Asada, K. F. MacDorman, H. Ishiguro, and Y. Kuniyoshi, "Cognitive developmental robotics as a new paradigm for the design of humanoid robots," *Robot. Auto. Syst.*, vol. 37, nos. 2–3, pp. 185–193, 2001.
[3] O. Scharenborg, "Reaching over the gap: A review of efforts to link human and automatic speech recognition research," *Speech Commun.*, vol. 49, no. 5, pp. 336–347, 2007.
[4] J. Blauert, *Spatial Hearing: The Psychophysics of Human Sound Localization*. Cambridge, MA, USA: MIT Press, 1997.
[5] E. Lopez-Poveda, A. Palmer, and R. Meddis, *The Neurophysiological Bases of Auditory Perception*. New York, NY, USA: Springer-Verlag, 2010.
[6] B. Moore, *An Introduction to the Psychology of Hearing*. Leiden, The Netherlands: Brill, 2013.
[7] T. Kanda, H. Ishiguro, M. Imai, and T. Ono, "Development and evaluation of interactive humanoid robots," *Proc. IEEE*, vol. 92, no. 11, pp. 1839–1850, Nov. 2004.
[8] J. Bauer, C. Weber, and S. Wermter, "A SOM-based model for multisensory integration in the superior colliculus," in *Proc. Int. Joint Conf. Neural Netw. (IJCNN)*, Jun. 2012, pp. 1–8.
[9] P. H. Smith, P. X. Joris, and T. C. T. Yin, "Projections of physiologically characterized spherical bushy cell axons from the cochlear nucleus of the cat: Evidence for delay lines to the medial superior olive," *J. Comparative Neurol.*, vol. 331, no. 2, pp. 245–260, 1993.
[10] P. X. Joris, P. H. Smith, and T. C. T. Yin, "Coincidence detection in the auditory system: 50 Years after Jeffress," *Neuron*, vol. 21, pp. 1235–1238, Dec. 1998.
[11] D. R. F. Irvine, V. N. Park, and L. McCormick, "Mechanisms underlying the sensitivity of neurons in the lateral superior olive to interaural intensity differences," *J. Neurophysiol.*, vol. 86, no. 6, pp. 2647–2666, 2001.
[12] S. M. Chase and E. D. Young, "Cues for sound localization are encoded in multiple aspects of spike trains in the inferior colliculus," *J. Neurophysiol.*, vol. 99, no. 4, pp. 1672–1682, 2008.
[13] L. A. Jeffress, "A place theory of sound localization," *J. Comparative Physiol. Psychol.*, vol. 41, no. 1, p. 35, 1948.
[14] J. Dávila-Chacón, S. Heinrich, J. Liu, and S. Wermter, "Biomimetic binaural sound source localisation with ego-noise cancellation," in *Proc. Int. Conf. Artif. Neural Netw. Mach. Learn. (ICANN)*, 2012, pp. 239–246.
[15] M. Cobos, A. Marti, and J. J. Lopez, "A modified SRP-PHAT functional for robust real-time sound source localization with scalable spatial sampling," *IEEE Signal Process. Lett.*, vol. 18, no. 1, pp. 71–74, Jan. 2011.
[16] L. O. Nunes *et al.*, "A steered-response power algorithm employing hierarchical search for acoustic source localization using microphone arrays," *IEEE Trans. Signal Process.*, vol. 62, no. 19, pp. 5171–5183, Oct. 2014.
[17] J.-M. Valin, F. Michaud, J. Rouat, and D. Letourneau, "Robust sound source localization using a microphone array on a mobile robot," in *Proc. IEEE Int. Conf. Intell. Robot. Syst. (IROS)*, vol. 2. Oct. 2003, pp. 1228–1233.
[18] Y. Tamai, Y. Sasaki, S. Kagami, and H. Mizoguchi, "Three ring microphone array for 3D sound localization and separation for mobile robot audition," in *Proc. IEEE Int. Conf. Intell. Robot. Syst. (IROS)*, Aug. 2005, pp. 4172–4177.
[19] C. L. Epifanio, "Acoustic daylight: Passive acoustic imaging using ambient noise," M.S. thesis, Univ. California, San Diego, CA, USA, 1997.
[20] H. Liu and M. Shen, "Continuous sound source localization based on microphone array for mobile robots," in *Proc. IEEE Int. Conf. Intell. Robot. Syst. (IROS)*, Oct. 2010, pp. 4332–4339.
[21] M. Ren and Y. X. Zou, "A novel multiple sparse source localization using triangular pyramid microphone array," *IEEE Signal Process. Lett.*, vol. 19, no. 2, pp. 83–86, Feb. 2012.
[22] D. Pavlidi, A. Griffin, M. Puigt, and A. Mouchtaris, "Real-time multiple sound source localization and counting using a circular microphone array," *IEEE Trans. Audio, Speech, Lang. Process.*, vol. 21, no. 10, pp. 2193–2206, Oct. 2013.
[23] J. C. Middlebrooks and D. M. Green, "Sound localization by human listeners," *Annu. Rev. Psychol.*, vol. 42, no. 1, pp. 135–159, Feb. 1991.
[24] K. Voutsas and J. Adamy, "A biologically inspired spiking neural network for sound source lateralization," *IEEE Trans. Neural Netw.*, vol. 18, no. 6, pp. 1785–1799, Nov. 2007.
[25] W. Maass, "Networks of spiking neurons: The third generation of neural network models," *Neural Netw.*, vol. 10, no. 9, pp. 1659–1671, 1997.
[26] W. Maass and C. M. Bishop, *Pulsed Neural Networks*. Cambridge, MA, USA: MIT Press, 2001.
[27] T. Rodemann, M. Heckmann, F. Joublin, C. Goerick, and B. Scholling, "Real-time sound localization with a binaural head-system using a biologically-inspired cue-triple mapping," in *Proc. IEEE Int. Conf. Intell. Robot. Syst. (IROS)*, Oct. 2006, pp. 860–865.
[28] V. Willert, J. Eggert, J. Adamy, R. Stahl, and E. Korner, "A probabilistic model for binaural sound localization," *IEEE Trans. Syst. Man, Cybern. B, Cybern.*, vol. 36, no. 5, pp. 982–994, Oct. 2006.
[29] J. Nix and V. Hohmann, "Sound source localization in real sound fields based on empirical statistics of interaural parameters," *J. Acoust. Soc. Amer.*, vol. 119, pp. 463–479, Jan. 2006.
[30] J. Liu, D. Perez-Gonzalez, A. Rees, H. Erwin, and S. Wermter, "A biologically inspired spiking neural network model of the auditory midbrain for sound source localisation," *Neurocomputing*, vol. 74, nos. 1–3, pp. 129–139, 2010.
[31] D. Gouaillier *et al.*, "Mechatronic design of NAO humanoid," in *Proc. IEEE Int. Conf. Robot. Autom. (ICRA)*, May 2009, pp. 769–774.
[32] J. Dávila-Chacón, S. Magg, J. Liu, and S. Wermter, "Neural and statistical processing of spatial cues for sound source localisation," in *Proc. IEEE Int. Joint Conf. Neural Netw. (IJCNN)*, Aug. 2013, pp. 1–8.
[33] R. Beira *et al.*, "Design of the robot-cub (iCub) head," in *Proc. IEEE Int. Conf. Robot. Autom. (ICRA)*, May 2006, pp. 94–100.
[34] M. Slaney, "An efficient implementation of the Patterson–Holdsworth auditory filter bank," Perception Group, Apple Comput., Cupertino, CA, USA, Tech. Rep. 35, 1993.
[35] A. Marti, M. Cobos, and J. J. Lopez, "Automatic speech recognition in cocktail-party situations: A specific training for separated speech," *J. Acoust. Soc. Amer.*, vol. 131, no. 2, pp. 1529–1535, 2012.

[36] F. Asano, M. Goto, K. Itou, and H. Asoh, "Real-time sound source localization and separation system and its application to automatic speech recognition," in *Proc. INTERSPEECH*, 2001, pp. 1013–1016.

[37] M. Fréchette, D. Létourneau, J.-M. Valin, and F. Michaud, "Integration of sound source localization and separation to improve dialogue management on a robot," in *Proc. IEEE Int. Conf. Intell. Robot. Syst. (IROS)*, Oct. 2012, pp. 2358–2363.

[38] C.-Q. Li, F. Wu, S.-J. Dai, L.-X. Sun, H. Huang, and L.-Y. Sun, "A novel method of binaural sound localization based on dominant frequency separation," in *Proc. IEEE Int. Congr. Image Signal Process. (CISP)*, Oct. 2009, pp. 1–4.

[39] A. Deleforge and R. Horaud, "The cocktail party robot: Sound source separation and localisation with an active binaural head," in *Proc. Int. Conf. Human-Robot Interact.*, 2012, pp. 431–438.

[40] J. Woodruff and D. Wang, "Binaural detection, localization, and segregation in reverberant environments based on joint pitch and azimuth cues," *IEEE Trans. Audio, Speech, Lang. Process.*, vol. 21, no. 4, pp. 806–815, Apr. 2013.

[41] M. Wilson, "Six views of embodied cognition," *Psychonomic Bull. Rev.*, vol. 9, no. 4, pp. 625–636, 2002.

[42] G. Metta, G. Sandini, D. Vernon, L. Natale, and F. Nori, "The iCub humanoid robot: An open platform for research in embodied cognition," in *Proc. ACM 8th Workshop Perform. Metrics Intell. Syst.*, 2008, pp. 50–56.

[43] J. Twiefel, T. Baumann, S. Heinrich, and S. Wermter, "Improving domain-independent cloud-based speech recognition with domain-dependent phonetic post-processing," in *Proc. AAAI*, 2014, pp. 1529–1536.

[44] J. Bauer, J. Dávila-Chacón, E. Strahl, and S. Wermter, "Smoke and mirrors—Virtual realities for sensor fusion experiments in biomimetic robotics," in *Proc. IEEE Int. Conf. Multisensor Fusion Integr. Intell. Syst. (MFI)*, Sep. 2012, pp. 114–119.

[45] R. Meddis, E. Lopez-Poveda, R. R. Fay, and A. N. Popper, *Computational Models of the Auditory System*, vol. 35. New York, NY, USA: Springer, 2010.

[46] J. Liu, H. Erwin, S. Wermter, and M. Elsaid, "A biologically inspired spiking neural network for sound localisation by the inferior colliculus," in *Proc. Int. Conf. Artif. Neural Netw. (ICANN)*, 2008, pp. 396–405.

[47] J. S. Garofolo, L. F. Lamel, W. M. Fisher, J. G. Fiscus, D. S. Pallett, and N. L. Dahlgren, "DARPA TIMIT: Acoustic-phonetic continous speech corpus CD-ROM. NIST speech disc 1-1.1," Defense Adv. Res. Projects Agency, Inf. Sci. Technol. Office, Gaithersburg, MD, USA, Tech. Rep. 4930, 1993.

[48] S. Salb and P. Duhr, "Comparison between Soundman OKM II Studio Classic and Neumann Dummy Head KU81i in technical and timbral aspects," SAE Inst., Univ. Middlesex, London, U.K., Tech. Rep. RA-303, 2009. [Online]. Available: http://dev.soundman.de/wp-content/uploads/download/thesis-EN.pdf

[49] S. Heinrich and S. Wermter, "Towards robust speech recognition for human-robot interaction," in *Proc. IROS Workshop Cognit. Neurosci. Robot. (CNR)*, 2011, pp. 29–34.

[50] J. Schalkwyk *et al.*, "'Your word is my command': Google search by voice: A case study," in *Proc. Adv. Speech Recognit.*, 2010, pp. 61–90.

[51] W. Walker *et al.*, "Sphinx-4: A flexible open source framework for speech recognition," Menlo Park, CA, USA, Tech. Rep. 2004-139, 2004.

[52] A. Rubruck *et al.*, "CoCoCo, coffee collecting companion," in *Proc. 8th AAAI Video Competition 28th Conf. Artif. Intell. (AAAI)*, Québec, CA, USA, 2014. [Online]. Available: [Online]. Available: https://www2.informatik.uni-hamburg.de/wtm/publications/2014/RAYHWKYTSBDHW14a/

[53] M. Bisani and H. Ney, "Joint-sequence models for grapheme-to-phoneme conversion," *Speech Commun.*, vol. 50, no. 5, pp. 434–451, 2008.

[54] V. I. Levenshtein, "Binary codes capable of correcting deletions, insertions and reversals," *Sov. Phys.—Dokl.*, vol. 10, pp. 707–710, Feb. 1966.

[55] J. Liu and G.-Z. Yang, "Robust speech recognition in reverberant environments by using an optimal synthetic room impulse response model," *Speech Commun.*, vol. 67, pp. 65–77, Mar. 2014.

[56] Y. Guo, X. Wang, C. Wu, Q. Fu, N. Ma, and G. J. Brown, "A robust dual-microphone speech source localization algorithm for reverberant environments," in *Proc. INTERSPEECH*, 2016, pp. 3354–3358.

[57] X. Zhang and D. Wang, "Deep learning based binaural speech separation in reverberant environments," *IEEE/ACM Trans. Audio, Speech, Language Process.*, vol. 25, no. 5, pp. 1075–1084, May 2017.

[58] J. Bauer and S. Wermter, "Self-organized neural learning of statistical inference from high-dimensional data," in *Proc. 23rd Int. Joint Conf. Artif. Intell. (IJCAI)*, 2013, pp. 1226–1232.

[59] E. M. Z. Golumbic *et al.*, "Mechanisms underlying selective neuronal tracking of attended speech at a 'cocktail party,'" *Neuron*, vol. 77, no. 5, pp. 980–991, 2013.

[60] P. Lakatos, G. Musacchia, M. N. O'Connel, A. Y. Falchier, D. C. Javitt, and C. E. Schroeder, "The spectrotemporal filter mechanism of auditory selective attention," *Neuron*, vol. 77, no. 4, pp. 750–761, 2013.

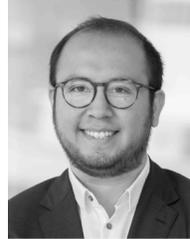

**Jorge Dávila-Chacón** received the B.Eng. double degree in mechanics and electricity from the Benemérita Universidad Autónoma de Puebla, Puebla, Mexico, and the M.Sc. degree in artificial intelligence from the University of Groningen, Groningen, The Netherlands. He is currently pursuing the Ph.D. degree in neural computation from the University of Hamburg, Hamburg, Germany.

He was with the Stem Cell and Brain Research Institute, CNRS, Lyon, France, as a Research Intern. He participated in several tournaments of RoboCup@Home, the largest international competition on domestic-service robots. He is a Founder of Heldenkombinat Technologies GmbH, Hamburg, where he is involved in designing AI solutions for the industry.

Mr. Jorge has served as Invited Reviewer for the *Journal of Computer Speech and Language*. He was one of the organizing committee chairs for ICANN 2014.

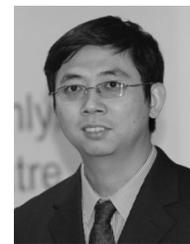

**Jindong Liu** (M'03) received the Ph.D. degree from the University of Essex, Colchester, U.K., with a focus on biologically inspired autonomous robotic fish.

From 2008 to 2010, he was with the University of Sunderland, Sunderland, U.K., in collaboration with the University of Newcastle, Newcastle, Australia, where he is involved in the development of a computational mammalian auditory system applied to the sound perception on mobile robotics. In 2010, he was with the Imperial College London, London, U.K., where he is a Research Fellow with the Hamlyn Centre for Robotic Surgery. He successfully built the first autonomous robotic fish. He has authored or co-authored articles in *Neurocomputing*, the *Journal of Bionic Engineering*, the *Journal of Neural Network World*, and the *International Journal of Automation and Computing*. His current research interests include biologically inspired mobile robotics, mainly including natural human–robot interaction, biomimetic robotic fish, and compliant manipulator for healthcare and surgery robotics.

Dr. Liu is a reviewer for conferences and journals of the IEEE and Springer, including the IEEE TRANSACTIONS OF NEURAL NETWORK AND LEARNING SYSTEMS, IROS, and ICRA. He was a recipient of the Best Poster Award in the 9th International Conference of Body Sensor Network in 2012.

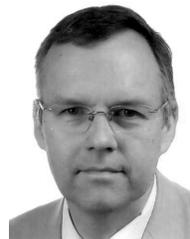

**Stefan Wermter** received the M.Sc. degree in computer science from the University of Massachusetts, MA, USA, and the Ph.D. and Habilitation degrees in computer science from the University of Hamburg, Hamburg, Germany.

He is a Full Professor with the University of Hamburg and the Head of the Knowledge Technology Group, University of Hamburg. He has been a Research Scientist with the International Computer Science Institute in Berkeley, Berkeley, CA, USA, before leading the Chair in Intelligent Systems at the University of Sunderland, Sunderland, U.K. His current research interests include neural networks, hybrid systems, cognitive neuroscience, cognitive robotics, and natural language processing.

Dr. Wermter was a General Chair for ICANN 2014, on the board of the European Neural Network Society. He is an Associate Editor of the journals *Connection Science*, the *International Journal for Hybrid Intelligent Systems*', the IEEE TRANSACTIONS ON NEURAL NETWORKS AND LEARNING SYSTEMS. He is on the Editorial Board of the journals *Cognitive Systems Research* and the *Journal of Computational Intelligence*.